\documentstyle[preprint,aps]{revtex}
\begin{document}
\draft

\title{Dynamic   spin susceptibility in the $t-J$ model}
\author{ G. Jackeli$^{a,b}$, N.M. Plakida$^{a}$}
\address{
$^{a}$Joint Institute for Nuclear Research, Dubna, Moscow region,
 141980, Russia \\
$^{b}$ Department of Solid State Physics,
Tbilisi State University, Tbilisi, Georgia}

\maketitle
\begin{abstract}\widetext
Dynamic spin susceptibility
is calculated for the $t-J$ model in the paramagnetic phase  by applying
the memory function method in terms of the Hubbard operators.
A self-consistent system of equations for the memory function is obtained
within the mode coupling approximation.
Both itinerant hole excitations and localized spin fluctuations  give
contributions to the memory function. Spin dynamics have a diffusive
character in the hydrodynamic limit; spin-wave-like excitations are regained
in the high-frequency region.
\end{abstract}
\narrowtext

\section{Introduction}
While the reference antiferromagnetic (AFM) insulator compounds
of high-$T_{c}$ superconductors (HTSC's)  are well understood in
terms of 2-dimensional (2D) isotropic Heisenberg model the nature
of anomalous spin-dynamics in the doped samples still requires
proper understanding ~\cite{kam}.

One of the simplest models invoked to describe the HTSC's is the $t-J$
model, which contains the essential physics of $CuO_{2}$ planes in the
superconducting cuprates. The $t-J$
(or its extension $t-t^{\prime}-J$) model is the low energy effective model
obtained from the Hubbard model by projecting out doubly occupied sites.
As a result the $t-J$ model is formulated in terms of the so called Hubbard
operators (HO's), which are neither Fermi nor Bose operators.
This particularity makes it difficult to treat the $t-J$ model within
the conventional field-theory methods.

The various approaches, e.g., slave-boson \cite{tan,ste} or
slave-fermion \cite{hed,kan,aue} methods and diagrammatic technique
for HO's  \cite{izu}, have been used to study the spin dynamics within the
$t-J$ model. In the slave-field approaches the local constraint is usually
replaced by the global one restricting the validity of the approach.
Whereas in any approximation formulated in terms of HO's the constraint of
no doubly occupancy can be rigorously preserved.

Recently in
Ref.~\cite{onu} the diagrammatic technique for HO's
has been used to calculate the spin susceptibility
within the Larkin equation \cite{lar}
for the $t-J$ model.
However, in the particular case $J=0$
(corresponding to $U\rightarrow \infty$ limit of the Hubbard model)
the contribution from the irreducible part in
the denominator of Larkin equation (see Eq. (9) of Ref.~\cite{onu})
vanishes, which indicates that the so called kinematical interaction
is not properly taken into account.

In the present paper we study the dynamic spin
susceptibility for the $t-J$ model by applying memory function technique
\cite{for} in term of HO's which has been applied recently
for calculation of the optical conductivity for this model
~\cite{pla}. We show that there exist two different in nature
contributions to the memory function. The first one is due to the kinematical
interaction and comes from the particle-hole excitations in
the itinerant hole subsystem.
While the second one comes from the localized spin fluctuations
due to the Heisenberg interaction.
The existence of these two contributions
explicitly shows that there is a competition between itinerant and localized
magnetism as it has been pointed out in Refs. ~\cite{izu,onu} and
observed experimentally (see, e. g. Ref. ~\cite{aep}).

It is found that the low energy ($\omega\rightarrow 0$) spin dynamics
has a diffusive character. While in the high energy limit
($\omega\rightarrow \infty$ ) the spin-wave-like excitations are
regained. The mean-filed-like (MFL) expression obtained earlier ~\cite{shi}
by Kondo and Yamaji's (KY's) theory ~\cite{kon} is recovered in this limit.

The paper is organized as follows. In the next section we formulate
the $t-J$ model in terms of HO's. In Sec. 3  the general formalism of the
memory function approach is presented and within
the mode coupling approximation
the memory function is calculated.
The self-consistency of the presented approach is discussed in Sec. 4.
Sec. 5 summarizes our main results. In the Appendix the expression for
static spin susceptibility is derived.

\section{ The model}
The Hamiltonian of the $t-J$ model reads
\begin{equation}
H=H_{t}+H_{J}=\sum_{i,j,\sigma}t_{ij}X_{i}^{\sigma 0}X_{j}^{0\sigma}+
\sum_{i,j}J_{ij}\Bigl\{{\bf S}_{i}{\bf S}_{j}-\frac{1}{4}n_{i}n_{j}\Bigr\}.
\label{m1}
\end{equation}
The first term in the right hand-side of Eq. (\ref{m1})
describes an electron
hopping between the nearest ($t_{ij}=t$) and next-nearest
($t_{ij}=t^{\prime}$) neighbor lattice sites. The second term
describes the exchange interaction of localized
spins ${\bf S}_{i}$ between the
nearest neighbor sites ($J_{ij}=J$).
The HO's in Eq. (\ref{m1}) are defined as
\begin{equation}
X_{i}^{\alpha\beta}=|i,\alpha\rangle\langle i,\beta|
\label{m2}
\end{equation}
for three possible states at the lattice site $i$
\begin{equation}
|i,\alpha\rangle=|i,0\rangle,\;\;\;|i,\sigma\rangle,
\label{m3}
\end{equation}
for an empty site and for a singly occupied site by electron
with spin $\sigma/2$ ($\sigma=\pm1$).  In the $t-J$ model only singly occupied sites are
retained and the completeness relation for the HO's reads as
\begin{equation}
X_{i}^{00}+\sum_{\sigma}X_{i}^{\sigma\sigma}=1.
\label{m4}
\end{equation}
The spin and density operators in Eq. (\ref{m1}) are expressed by
HO's as
\begin{equation}
S_{i}^{\sigma}=X_{i}^{\sigma\bar{\sigma}},\;\;\;
S_{i}^{z}=\frac{1}{2}\sum_{\sigma}\sigma X_{i}^{\sigma\sigma},\;\;\;
n_{i}=\sum_{\sigma}X_{i}^{\sigma\sigma}.
\label{m5}
\end{equation}
where $\bar{\sigma}=-\sigma$. The HO's obey the following multiplication
rules
\begin{equation}
X_{i}^{\alpha\beta}X_{j}^{\gamma\delta}=\delta_{ij}\delta_{\beta\gamma}
X_{i}^{\alpha\delta}
\label{m6}
\end{equation}
and commutation relations
\begin{equation}
\left[X_{i}^{\alpha\beta}X_{j}^{\gamma\delta}\right]_{\pm}=
\delta_{ij}\left(\delta_{\beta\gamma}X_{i}^{\alpha\delta}\pm
\delta_{\delta\alpha}X_{i}^{\gamma\beta}\right).
\label{m7}
\end{equation}
In Eq.(\ref{m7}) the upper sign stands for the case when both
HO's are Fermi-like ones (as, e. g., $X_{i}^{0\sigma}$).
The spin and density operators (\ref{m5}) are Bose-like and for them
the lower sign in Eq.(\ref{m7}) should be taken.
The HO's are neither Fermi nor Bose operators, the are projected
operators. These unconventional commutation relations (\ref{m7})
makes impossible to treat the model within the
conventional diagrammatic technique. To use
the latter one needs to introduce slave particles with the
constraint of no doubly occupancy. While in the treatment within
the HO's the constraint is automatically fulfilled.
Therefore we treat the problem in terms of HO's and use
the memory function formalism to determine the dynamic spin
susceptibility.

\section{Memory Function}
\subsection{General Formalism}

The dynamic spin susceptibility is defined as
\begin{equation}
\chi_{\bf q}(\omega)=-\langle\langle
S^{+}_{\bf q}|S^{-}_{\bf q}\rangle\rangle_{\omega}=-
\sum_{{\bf R}_{ij}}e^{-i{\bf q}{\bf R}_{ij}}
\langle\langle
S^{+}_{i}|S^{-}_{j}\rangle\rangle_{\omega}
,
\label{g1}
\end{equation}
where ${\bf R}_{ij}={\bf R}_{i}-{\bf R}_{j}$ and
$\langle\langle A|B\rangle\rangle_{\omega}$ denotes the Fourier transformed two-time retarded
commutator Green function (GF) ~\cite{zub,tse}
\begin{equation}
\langle\langle A|B\rangle\rangle_{\omega}=-i \int_{0}^{\infty}dt e^{i\omega t}
\langle[A(t), B]\rangle,
\label{g2}
\end{equation}
where $\rm{Im}\omega >0$, $ A(t)=\exp (i Ht) A\exp (-i Ht)$,
and $\langle AB \rangle$ denotes the equilibrium statistical average.

In the paramagnetic state
with zero sublattice magnetization
%where only short-range AFM correlations persist,
 an average of the
commutator $\langle[S^{+}_{i},S^{-}_{j}]\rangle
=2\delta_{ij}\langle S_{i}^{z}\rangle$ equals zero.
Since just this quantity enters as an initial condition $(t=0)$ in the
equation of motion for the GF (\ref{g1}), it is more convenient to
construct the self-consistent equation for the Kubo-Mori
relaxation function (see, e.g., Ref.~\cite{for})
\begin{equation}
(( A | B ))_{\omega} = - i \int_{0}^{\infty} dt  e^{i
   \omega t}  (A(t), B),
\label{g3}
\end{equation}
where $(A(t), B)$ is the Kubo-Mori scalar product defined as
\begin{equation}
(A(t), B) =\int_{0}^{\beta}d\lambda \langle A(t-i\lambda) B \rangle,
\label{g4}
\end{equation}
where $\beta=1/T$ $(h=k_{B}=1)$.

The GFs (\ref{g2}) and (\ref{g3}) are coupled by the equation
\begin{equation}
\omega  (( A | B ))_{\omega} =
\langle\langle A | B \rangle\rangle_{\omega} -
\langle\langle A | B \rangle\rangle_{\omega =0} .
\end{equation}
There are also following useful relations which can be obtained
from the definitions (\ref{g2})-(\ref{g4})
\begin{equation}
(( i\dot{A}| B ))_{\omega}
=(( A | -i\dot{ B} ))_{\omega}
=\langle\langle A | B \rangle\rangle_{\omega},
\label{g6}
\end{equation}
\begin{equation}
( i \dot{A} , B ) =  ( A , - i\dot{ B} )
=\langle [ A , B ]\rangle,
\label{g7}
\end{equation}
\begin{equation}
(A , B )= - \langle\langle A | B \rangle\rangle_{\omega =0},
\label{g8}
\end{equation}
where  $ i \dot{A} = i dA/dt = [A,H] $.

By using the above formulas for the dynamic spin susceptibility we obtain
\begin{equation}
\chi_{\bf q}(\omega)=\chi_{\bf q}-\omega\Phi_{q}(\omega),
\label{g9}
\end{equation}
where $\chi_{\bf q}\equiv\chi_{\bf q}(0)$ is the static spin
susceptibility and
$\Phi_{\bf q}(\omega)\equiv((S^{+}_{\bf q}|S^{-}_{\bf q}))_{\omega}$.
To calculate the spin-spin relaxation function $\Phi_{\bf q}(\omega)$
it is convenient to employ the memory function approach of Mori
(see, e.g. Ref.~\cite{for}).
We define the memory function $M({\bf q},\omega)$ by the equation
\begin{equation}
\Phi_{\bf q}(\omega)=\frac{\chi_{\bf q}}
{\omega - M({\bf q},\omega)/\chi_{\bf q}}
\label{g10}
\end{equation}

To calculate the memory function we use the equation of motion for the
relaxation function (\ref{g3})
\begin{equation}
\omega((S_{\bf q}^{+}|S_{\bf q}^{-}))_{\omega}=\chi_{\bf q}+
((i\dot{S}_{\bf q}^{+}|S_{\bf q}^{-}))_{\omega},
\label{g12}
\end{equation}
and similarly  $((i\dot{S}_{\bf q}^{+}|S_{\bf q}^{-}))_{\omega}$
obey the following equation of motion
\begin{equation}
\omega((i\dot{S}_{\bf q}^{+}|S_{\bf q}^{-}))_{\omega}=
(i\dot{S}_{\bf q}^{+},S_{\bf q}^{-})+
((i\dot{S}_{\bf q}^{+}|-i\dot{S}_{\bf q}^{-}))_{\omega}.
\label{g13}
\end{equation}
From Eq.(\ref{g7}) we have
$(i\dot{S}_{\bf q}^{+},S_{\bf q}^{-})=
\langle[S_{\bf q}^{+},S_{\bf q}^{-}]\rangle=
2/\sqrt{N}\langle S_{\bf q}^{z}\rangle\delta_{{\bf q}0}$
which is zero in the paramagnetic phase, that results
\begin{equation}
\omega\Phi_{\bf q}(\omega)=\chi_{\bf q}+\frac{1}{\omega}
((i\dot{S}_{\bf q}^{+}|-i\dot{S}_{\bf q}^{-}))_{\omega}.
\label{g14}
\end{equation}

By introducing the zero order GF
$\Phi_{\bf q}^{0}(\omega)=\chi_{\bf q}/\omega$ we rewrite Eq. (\ref{g14}) as
follows
\begin{equation}
\Phi_{\bf q}(\omega)=\Phi_{\bf q}^{0}(\omega)+
\Phi_{\bf q}^{0}(\omega)T_{\bf q}(\omega)
\Phi_{\bf q}^{0}(\omega)
\label{g15}
\end{equation}
where we have introduced the scattering matrix
\begin{equation}
T_{\bf q}(\omega)=\frac{1}{\chi_{\bf q}}
((i\dot{S}_{\bf q}^{+}|-i\dot{S}_{\bf q}^{-}))_{\omega}
\frac{1}{\chi_{\bf q}}.
\label{g16}
\end{equation}
By comparing (\ref{g16}) to the definition of the memory function
(\ref{g10}) we get the following relation between the scattering
matrix and the memory function
\begin{equation}
T_{\bf q}(\omega)=\frac{M({\bf q},\omega)}{\chi_{\bf q}^{2}}+
\frac{M({\bf q},\omega)}{\chi_{\bf q}^{2}}
\Phi_{\bf q}^{0}(\omega)T_{\bf q}(\omega).
\label{g17}
\end{equation}
A formal solution of the Eq. (\ref{g17}) by iteration shows that the
quantity $M({\bf q},\omega)/\chi_{\bf q}(\omega)$ is just the
irreducible part of the scattering matrix which has no parts connected
by the single zero order GF $\Phi_{\bf q}^{0}(\omega)$:
\begin{equation}
M({\bf q},\omega)=\chi_{\bf q}^{2}T_{\bf q}^{irr}(\omega)=
((i\dot{S}_{\bf q}^{+}|-i\dot{S}_{\bf q}^{-}))_{\omega}^{irr}
\label{g18}
\end{equation}

Finally, the dynamic spin susceptibility in terms
of the memory function can be written as
\begin{equation}
\chi_{\bf q}(\omega)=-\chi_{\bf q}\frac{M({\bf q},\omega)/\chi_{\bf q}}
{\omega-M({\bf q},\omega)/\chi_{\bf q}}.
\label{g19}
\end{equation}
\subsection{Mode-coupling approximation}

First we express the memory function in terms of the irreducible
current-current time-dependent correlation function by using the spectral
representation for the GF

\begin{equation}
M({\bf q}, \omega)=\sum_{{\bf R}_{ij}}e^{-{\bf q}{\bf R}_{ij}}\frac{1}{2\pi}
\int_{-\infty}^{\infty}d\omega^{\prime}\frac{e^{\beta\omega^{\prime}}-1}
{\omega^{\prime}(\omega-\omega^{\prime}+i\eta)}\int_{-\infty}^{\infty}
dte^{-i\omega^{\prime}t}\langle J^{\dagger}_{i}(t)|J_{j} \rangle^{irr} ,
\label{g20}
\end{equation}
where the current operator in
the site representation is defined as  $J_{j}=i\dot{S}_{j}^{+}$. Current operator can be written as a
sum of two terms
\begin{equation}
J_{j}=J_{j}^{t}+J_{j}^{J}=[S_{j}^{+},H_{t}]+[S_{j}^{+},H_{J}]
\label{g21}
\end{equation}

In Eq.(\ref{g21}) the first term comes from the so called
kinematical interaction
which is due to the unconventional commutation relation for the HO's
operators (\ref{m7}). This term is proportional to the hopping integral and reads as
\begin{equation}
J_{j}^{t}=-\sum_{m}t_{jm}(X_{j}^{+0}X_{m}^{0-}-X_{m}^{+0}X_{j}^{0-}).
\label{g22}
\end{equation}

The second term in Eq.(\ref{g21}) comes from the exchange interaction between
localized spins and has the form
\begin{equation}
J_{j}^{J}=2\sum_{m}J_{jm}(S_{j}^{z}S_{m}^{+}-S_{m}^{z}S_{j}^{+}).
\label{g23}
\end{equation}

To calculate the irreducible time-dependent correlation function in
the right hand -side of Eq. (\ref{g20}) we employ the mode-coupling approximation
in terms of an independent propagation of the dressed
particle-hole and spin fluctuations
(see, e. g., G\"{o}tze et al.,~\cite{got}).
 This scheme is essentially equivalent to
the self-consistent Born approximation in which the vertex corrections
are neglected. The proposed scheme is defined by the following decoupling of the
time-dependent correlation functions:
\begin{equation}
\langle X_{m}^{-0}(t)X_{i}^{0+}(t)X_{j}^{+0}X_{l}^{0-}\rangle\simeq
\langle X_{m}^{-0}(t)X_{l}^{0-}\rangle
\langle X_{i}^{0+}(t)X_{j}^{+0}\rangle,
\label{g24}
\end{equation}
\begin{equation}
\langle S_{i}^{z}(t)S_{m}^{-}(t)S_{j}^{z}S_{l}^{+}\rangle\simeq
\langle S_{i}^{z}(t)S_{j}^{z}\rangle
\langle S_{m}^{-}(t)S_{l}^{+}\rangle.
\label{g25}
\end{equation}
The cross-correlations like $\langle J_{i}^{t}|(J^{J})_{j}^{+}\rangle$
are ignored within the proposed approximation and they will be omitted.

By using the above defined decoupling scheme (\ref{g24}) and (\ref{g25}) and the
spectral representation for the GF, the memory function (\ref{g20})
can be written as
\begin{equation}
M({\bf q},\omega)=M_{t}({\bf q},\omega)+M_{J}({\bf q},\omega),
\label{g26}
\end{equation}
where $M_{t}({\bf q},\omega)$ is the contribution from the itinerant hole
subsystem and reads as
\begin{equation}
M_{t}({\bf q},\omega)=
\frac{1}{N}\sum_{{\bf k}}^{}t^{2}_{{\bf k}{\bf q}}
\int\!\!\!\!\int\limits_{-\infty }^{\infty }
d\omega_{1}d\omega^{\prime}
\left[ n(\omega_{1}-\omega^{\prime})
-n(\omega_{1})\right]
\frac{A_{{\bf k}}(\omega_{1})A_{{\bf k}-{\bf q}}(\omega_{1}-\omega^{\prime})}
{\omega^{\prime}(\omega-\omega^{\prime}+i\eta)}
,
\label{g27}
\end{equation}
where $n(\omega)=(e^{\beta\omega}+1)^{-1}$,
$t_{{\bf k}{\bf q}}=t_{\bf k}-t_{{\bf k}-{\bf q}}$ with
$t_{\bf k}=z(t\gamma_{\bf q}+t^{\prime}\gamma_{\bf q}^{\prime})$,
$z=4$,
$\gamma_{\bf q}=1/2 [\cos q_{x}+\cos q_{y}]$ and
$\gamma_{\bf q}^{\prime}=\cos q_{x}\cos q_{y}$ for 2D square lattice
(the lattice constant is taken to be unity)
and
$A_{\bf k}(\omega)=-1/\pi\rm{Im}\langle\langle
X_{\bf q}^{\sigma 0}X_{\bf q}^{0\sigma}\rangle\rangle_{\omega}$ is
a hole particle spectral function which is spin independent in the
paramagnetic phase.

The second contribution $M_{J}({\bf q},\omega)$ in Eq.(\ref{g26})
comes from the localized spin subsystem and is given by
\begin{equation}
M_{J}({\bf q},\omega)=
\frac{2}{\pi^{2}N}\sum_{{\bf k}}^{}J^{2}_{{\bf k}{\bf q}}
\int\!\!\!\!\int\limits_{-\infty }^{\infty }
d\omega_{1}d\omega^{\prime}
\left[ N(\omega_{1}-\omega^{\prime})
-N(\omega_{1})\right]
\frac{\rm{Im}\chi_{{\bf k}}(\omega_{1})\rm{Im}\chi_{{\bf k}-{\bf q}}
(\omega_{1}-\omega^{\prime})}
{\omega^{\prime}(\omega-\omega^{\prime}+i\eta)}
,
\label{g28}
\end{equation}
where $N(\omega)=(e^{\beta\omega}-1)^{-1}$ and
$J_{{\bf k}{\bf q}}=J_{\bf k}-J_{{\bf k}-{\bf q}}$, and
$J_{\bf q}=zJ\gamma_{\bf q}$.
In obtaining (\ref{g28}) relation
$\langle\langle S_{\bf q}^{z}|S_{\bf q}^{z}\rangle\rangle_{\omega}=
1/2\langle\langle S_{\bf q}^{+}|S_{\bf q}^{-}\rangle\rangle_{\omega}$
which is valid in the rotationally invariant system has been used.

The real, $\rm{Re}M({\bf q},\omega)$, and imaginary,
 $\rm{Im}M({\bf q},\omega)$,
parts of the memory function are odd and
even functions of $\omega$ , respectively, and they are coupled by the
dispersion relation
\begin{equation}
\rm{ReM}({\bf q},\omega) = \frac{1}{\pi}\int_{-\infty}^{\infty} d\omega^{\prime}
\frac {\rm{Im}M({\bf q},\omega^{\prime})}{\omega^{\prime}-\omega}.
\label{g29}
\end{equation}
Therefore, only imaginary part of the memory function should be evaluated.
\subsection{Asymptotic behavior of $\chi_{\bf q}(\omega)$}
Now we examine asymptotic behavior of the dynamic spin susceptibility.
First we consider the hydrodynamic limit ${\bf q}\rightarrow 0$ and
$\omega\rightarrow 0$. In this limit, $\rm{Re}M({\bf q},\omega)$
being an odd function of $\omega$ vanishes while $\rm{Im}M({\bf q},\omega)$
remains finite. By using Eqs.(\ref{g27}) and ({\ref{g28}) we
can express it as $\rm{Im}M({\bf q},\omega)\simeq -Dq^{2}$
with $D=D_{t}+D_{J}$ where
\begin{eqnarray}
D_{t}=
\frac{\pi}{N}\sum_{{\bf k}}^{}
(\hat{\bf q}\nabla_{\bf k}t_{\bf k})^{2}
\lim\limits_{\omega\to 0}
\lim\limits_{{\bf q}\to 0}
\int\limits_{-\infty }^{\infty }
d\omega_{1}
 |n^{\prime}(\omega_{1})|
A_{{\bf k}}(\omega_{1})
A_{{\bf k}-{\bf q}}(\omega_{1}-\omega),\nonumber\\
D_{J}=
\frac{2}{\pi N}\sum_{{\bf k}}^{}
(\hat{\bf q}\nabla_{\bf k}J_{\bf k})^{2}
\lim\limits_{\omega\to 0}
\lim\limits_{{\bf q}\to 0}
\int\limits_{-\infty }^{\infty }
d\omega_{1}
|N^{\prime}(\omega_{1})|
\rm{Im}\chi_{{\bf k}}(\omega_{1})
\rm{Im}\chi_{{\bf k}-{\bf q}}(\omega_{1}-\omega),
\label{g30}
\end{eqnarray}
where $\hat{\bf q}={\bf q}/q$, $\nabla_{\bf k}=dt_{\bf k}/d{\bf k}$,
$n^{\prime}(\omega)=dn(\omega)/d\omega$, and
$N^{\prime}(\omega)=dN(\omega)/d\omega$.
Finally in the hydrodynamic limit the dynamic spin-susceptibility can be
expressed in the usual form (see Ref. ~\cite{for}) as
\begin{equation}
\chi_{\bf q}(\omega)=\chi_{0}\frac{i\tilde{D}q^{2}}{\omega+i\tilde{D}q^{2}}
\label{g31}
\end{equation}
where $\tilde{D}=D/\chi_{0}$ is the spin diffusion coefficient and $\chi_{0}$
is the static uniform susceptibility.

Unlike to the hydrodynamic limit, in the high energy limit,
$\omega\rightarrow \infty$ the dominant contribution to the memory
functions comes from the real part:
$M({\bf q},\omega)\simeq m_{\bf q}/\omega$ where $m_{\bf q}$ is the first
non-vanishing moment in $1/\omega$ expansion of the memory function
defined as
\begin{equation}
m_{\bf q}=-\frac{1}{\pi}\int\limits_{-\infty}^{\infty}d\omega\rm{Im}M({\bf q},\omega)=
\langle[i\dot{S}_{\bf q}^{+},S_{\bf q}^{+}]\rangle
\label{g32}
\end{equation}
Thus in the high energy limit dynamic susceptibility takes the form
\begin{equation}
\chi_{\bf q}(\omega)=-\chi_{\bf q}\frac{\omega_{\bf q}^{2}}
{\omega^{2}-\omega_{\bf q}^{2}}
\label{g33}
\end{equation}
where $\omega_{\bf q}^{2}=m_{\bf q}/\chi_{\bf q}$.
Let us note, that Eq. (\ref{g33}) with  expressions of $\chi_{\bf q}$
and $m_{\bf q}$ derived in the Appendix [see Eqs. (\ref{A16}) and
(\ref{A17})] reproduces the result for the spin susceptibility
obtained for the $t-J$ model in Ref.~\cite{shi}
by KY's theory ~\cite{kon} which is essentially
self-consistent MFL approximation.
\section{Self-consistency of the problem}
%\sectio{Results and discussion}
%\section{Model calculations}
The equations (\ref{g19}),(\ref{g27}), and (\ref{g28})
are the self-consistent integral equations for dynamic
spin susceptibility which is obtained by using
only mode coupling approximation.
These equations should be solved numerically by iteration
procedure. The static spin susceptibility at each iteration
step should be calculated by Eq. (\ref{A4}) with $M({\bf q},\omega)$
and $\chi_{\bf q}(\omega)$ from the preceding iteration.
However some ansatz for $\chi_{\bf q}(\omega)$ as the starting point
of iteration procedure should be defined. Moreover we need to know
the hole spectral function entering into Eq. (\ref{g27})
for the memory function.

According to well known results for hole spectral function
obtained within $t-J$ model~\cite{dag} $A_{\bf k}(\omega)$
can be modeled as
\begin{equation}
A_{\bf k}(\omega)=Z_{\bf k}\delta(\omega+\mu-\varepsilon_{\bf k})+
A^{inc}_{\bf k}(\omega),
\label{r1}
\end{equation}
where $Z_{\bf k}$ is the quasiparticle weight for the excitations with the
dispersion $\varepsilon_{\bf k}$ in a narrow band of the order $J$.
The second part $A_{\bf k}^{inc}(\omega)$
is due to the diffusive motion of holes in a broad band with bandwidth $2W$
(of order $8t$ for 2D square lattice). We model it as
\begin{equation}
A_{\bf k}^{inc}=N_{inc}\theta(W-|\omega+\mu|),
\label{r2}
\end{equation}
where $N_{inc}$ is the density of state for the incoherent continuum
and it is coupled to $Z_{\bf k}$ by the sum rule
\begin{equation}
\frac{1}{N}\sum_{\bf k}\int_{-\infty}^{\infty}d\omega A_{\bf k}(\omega)=
\langle X_{i}^{00}+X_{i}^{\sigma\sigma}\rangle=1-\frac{n}{2}.
\label{r3}
\end{equation}
By using Eq.(\ref{r1}) the contribution from the hole coherent motion
to the imaginary part of the memory function can be expressed as
\begin{equation}
\rm{Im}M_{t}^{c-c}({\bf q},\omega)=\frac{\pi}{N}
\sum_{\bf k}t_{{\bf k}{\bf q}}^{2}
Z_{\bf k}Z_{{\bf k}-{\bf q}}
\frac{n(\varepsilon_{{\bf k}})-n(\varepsilon_{{\bf k}-{\bf q}})}
{\omega}\delta(\varepsilon_{\bf k}-\varepsilon_{{\bf k}-{\bf q}}-\omega)
\label{r4}
\end{equation}

To evaluate the second term in the memory function (\ref{g28})
as the starting point for $\chi_{\bf q}(\omega)$ one can use
the MFL expression (\ref{g33}),  with $m_{\bf q}$
 and $\omega_{\bf q}$ defined by (\ref{A8}) and
(\ref{A16}), respectively. As a result we obtain
\begin{equation}
\rm{Im}M_{J}({\bf q},\omega)=\frac{\pi}{2N}
\sum_{\bf k}B_{{\bf k}{\bf q}}\{P_{{\bf k}{\bf q}}(\omega)+
P_{{\bf k}{\bf q}}(-\omega)\}
\label{r5}
\end{equation}
where
\begin{equation}
B_{{\bf k}{\bf q}}=J_{{\bf k}{\bf q}}^{2}\frac{m_{\bf k}m_{{\bf k}-{\bf q}}}
{\omega_{{\bf k}}\omega_{{\bf k}-{\bf q}}},
\label{r6}
\end{equation}
is an effective vertex function and
\begin{equation}
P_{{\bf k}{\bf q}}(\omega)=\Biggl\{\frac{N(\omega_{{\bf k}})-N(\omega_{{\bf k}-{\bf q}})}
{\omega}\delta(\omega_{\bf k}-\omega_{{\bf k}-{\bf q}}-\omega)-
\frac{1+N(\omega_{{\bf k}})+N(\omega_{{\bf k}-{\bf q}})}
{\omega}\delta(\omega_{\bf k}+\omega_{{\bf k}-{\bf q}}-\omega)
\Biggr\}.
\label{r7}
\end{equation}

Eqs.(\ref{r4}) and Eq.(\ref{r5}) can be considered as the first iteration
for the memory function. The hole parameters entering into expressions
(\ref{A8}) and (\ref{A16}) for $m_{\bf q}$  and
$\omega_{\bf q}$ can be calculated from the hole spectral function
(\ref{r1}). Whereas
$\chi_{1}$ and $\chi_{2}$ defined by Eqs. (\ref{A10}) and (\ref{A18})
should be evaluated self-consistently from the dynamic spin
susceptibility (\ref{g33}).
By using the above expressions  (\ref{r4}) and (\ref{r5})
for $\rm{Im}M({\bf q},\omega)$ and the dispersion relation
(\ref{g29}) the dynamic spin susceptibility within
the first iteration can be calculated from Eq.(\ref{g19}).

Using the obtained
results one can evaluate the spin fluctuation part of
the memory (\ref{g28}) and the static spin susceptibility
(\ref{A4})-(\ref{A6}) for the next iteration procedure.
The iteration procedure should be continued until
the convergency will be reached.

%We hope that the obtained self-consistent
%mean field solution for $\chi_{\bf q}(\omega)$ is
%a good ansatz for above discussed integral equations.

\section{Summary}

To summarize, based on the $t-J$ model and memory function approach we
have derived a general representation for dynamic spin-susceptibility
(\ref{g19}) in terms of the memory function (\ref{g20}).
Our approach is formulated in terms of HO's and therefore the constraint
of no doubly occupancy is rigorously preserved.
The memory function is calculated by using the equation of motion method
for two-time retarded GF's~\cite{zub} within the mode coupling approximation
(\ref{g26}-\ref{g28}).
The two contributions to the memory function is obtained. The first
one (\ref{g27}) comes from the itinerant hole subsystem
and is due to the kinematical
interaction. The second one (\ref{g28}) comes from the localized
interacting spin subsystem.
In the limit of small concentration of doped holes the latter one gives
the main contribution which describes spin dynamics characteristic for
the Heisenberg model. Whereas in the opposite limit
of large hole concentration particle-hole excitations characteristic
to the itinerant magnetism give the main contribution to spin
dynamics. We have shown that in the paramagnetic phase there are two regimes
in the spin dynamics. In the hydrodynamic limit
($q\rightarrow 0$, $ \omega\rightarrow 0$) the spin susceptibility
(\ref{g31}) describes diffusion spin dynamics with the diffusion
coefficient (\ref{g30}), which has essentially two contributions.
While in the high-frequency limit
($\omega\rightarrow \infty$) spin-wave-like excitations described by
Eq. (\ref{g33}) are observed. Their dispersion, Eq. (\ref{A16}),
obtained in the mode coupling like approximation for the equal time
correlation function
(\ref{A14})
recovers the earlier MFL result~\cite{shi}
obtained within the KY's ~\cite{kon}theory.

To compare our results with that obtained by diagrammatic methods we would
like to point out that our approach, based on the general representation
for the spin susceptibility (\ref{g19}), is equivalent to summation
of infinite series of diagrams generated by the memory function (\ref{g20}).
The latter one, being calculated in the mode coupling approximation
,Eqs. (\ref{g24}) and (\ref{g25}), can be
schematically represented by two loop-diagrams:
the first one of order $t^{2}$ due to the particle-hole loop and
the second one of order $J^{2}$ due to the spin fluctuation loop.
In Ref.~\cite{onu} all the contributions in the denominator
of the Larkin equation are proportional to $J$ and therefore
disappears in the limit $J=0$ ($U\rightarrow \infty$).
While in our approach contribution due to the first loop
(for which the kinematical interaction is responsible)
remains. Whereas, in Ref. \cite{izu} the
spin fluctuation contribution given by our second loop is neglected while
several other diagrams
beyond our simple one-loop diagram
due to the kinematical interaction are
taken into account.

At present time it is difficult to justify any of discussed scheme,
including our mode coupling approximation. To check the validity of our
approximation one has to solve numerically self-consistent equations
and compare the obtained results with the
experimental data. This will be done in a forthcoming publication.

%\newpage
%\acknowledgments
%\begin{center}
%{\bf Acknowledgments}

\acknowledgments

Financial supports by the Russian
Foundation for Fundamental Researches Grant No 96--02--17527, and
the INTAS--RFBR Program Grant No 95--591 are acknowledged.
 Partial financial supports by NREL, Subcontract  No. AAX-6-16763-01,
by the Russian State Program ``High-Temperature Superconductivity'',
 Grant No 95056, and  the Heisen\-berg-Landau Program of BLTP, JINR,
 are also acknowledged.

\appendix
\section*{}
In this Appendix we evaluate the static spin
susceptibility $\chi_{\bf q}$ following Tserkovnikov ~\cite{tse1}.
For this purpose, by using Eqs.(\ref{g10}), (\ref{g13}), and
(\ref{g14}) we can rewrite the memory function or the irreducible part of
the current-current correlation function in the following way
proposed by Tserkovnikov~\cite{tse}:
\begin{equation}
((i\dot{S}_{\bf q}^{+}|-i\dot{S}_{\bf q}^{-}))_{\omega}^{irr}=
((i\dot{S}_{\bf q}^{+}|-i\dot{S}_{\bf q}^{-}))_{\omega}-
((i\dot{S}_{\bf q}^{+}|S_{\bf q}^{-}))_{\omega}
((S_{\bf q}^{+}|S_{\bf q}^{-}))_{\omega}^{-1}
((S_{\bf q}^{+}|-i\dot{S}_{\bf q}^{-}))_{\omega}.
\label{A1}
\end{equation}
Likewise we define the irreducible part of the force-force
correlation function as
\begin{equation}
((i^{2}\ddot{S}_{\bf q}^{+}|i^{2}\ddot{S}_{\bf q}^{-}))_{\omega}^{irr}=
((i^{2}\ddot{S}_{\bf q}^{+}|i^{2}\ddot{S}_{\bf q}^{-}))_{\omega}-
((i^{2}\ddot{S}_{\bf q}^{+}|S_{\bf q}^{-}))_{\omega}
((S_{\bf q}^{+}|S_{\bf q}^{-}))_{\omega}^{-1}
((S_{\bf q}^{+}|i^{2}\ddot{S}_{\bf q}^{-}))_{\omega},
\label{A2}
\end{equation}
and consequently the irreducible part of equal time force-force correlation
function can be written as
\begin{equation}
(i^{2}\ddot{S}_{\bf q}^{+},i^{2}\ddot{S}_{\bf q}^{-})^{irr}=
(i^{2}\ddot{S}_{\bf q}^{+},i^{2}\ddot{S}_{\bf q}^{-})-
(i^{2}\ddot{S}_{\bf q}^{+},S_{\bf q}^{-})
(S_{\bf q}^{+},S_{\bf q}^{-})^{-1}
(S_{\bf q}^{+},i^{2}\ddot{S}_{\bf q}^{-}).
\label{A3}
\end{equation}

By using Eq. (\ref{A3}) and identities (\ref{g6})-(\ref{g8})  the static
spin susceptibility can be expressed as
\begin{equation}
\chi_{\bf q}=\frac{\langle[i\dot{S}_{\bf q}^{+},S_{\bf q}^{-}]\rangle^{2}}
{(i^{2}\ddot{S}_{\bf q}^{+},i^{2}\ddot{S}_{\bf q}^{-})-
(i^{2}\ddot{S}_{\bf q}^{+},i^{2}\ddot{S}_{\bf q}^{-})^{irr}}.
\label{A4}
\end{equation}
The first term in the denominator can be written as the third
moment of the dynamic spin susceptibility
\begin{equation}
(i^{2}\ddot{S}_{\bf q}^{+},i^{2}\ddot{S}_{\bf q}^{-})=
\langle [i^{2}\ddot{S}_{\bf q}^{+},-i\dot{S}_{\bf q}^{-}]\rangle=
\frac{1}{\pi}\int\limits_{-\infty}^{\infty}d\omega\;\omega^{3}
\rm{Im}\chi_{\bf q}(\omega)
\label{A5}
\end{equation}
and the latter one is equal to the second non-vanishing moment of the
memory function
\begin{equation}
(i^{2}\ddot{S}_{\bf q}^{+},i^{2}\ddot{S}_{\bf q}^{-})^{irr}
=-\frac{1}{\pi}\int\limits_{-\infty}^{\infty}d\omega\;\omega^{2}
\rm{Im}M({\bf q},\omega).
\label{A6}
\end{equation}
The expression (\ref{A4}) is an exact representation for
the static spin susceptibility.

However, to derive
an approximate expression for $\chi_{\bf q}$ we start from the following
identity
\begin{equation}
\langle [i\dot{S}_{\bf q}^{+},S_{\bf q}]\rangle
=(i^{2}\ddot{S}_{\bf q}^{+},S_{\bf q}^{-}).
\label{A7}
\end{equation}
We evaluate the left hand-side of Eq. (\ref{A7}) by using the commutation
relation for HO's, that results
\begin{equation}
m_{\bf q}=\langle [i\dot{S}^{+}_{\bf q}|S_{\bf q}^{-}]\rangle=
4zJ(1-\gamma_{\bf q})\Bigl\{
\frac{t}{2J}n_{1}+\frac{t^{\prime}}{2J}n_{1}^{\prime}\lambda_{\bf q}
-\chi_{1}\Bigr\},
\label{A8}
\end{equation}
with the following notations
\begin{equation}
n_{1}=\frac{1}{N}\sum_{\bf q}\gamma_{\bf q}n_{\bf q},\;\;\;
n^{\prime}=\frac{1}{N}\sum_{\bf q}\gamma_{\bf q}^{\prime}n_{\bf q},\;\;\;
n_{\bf q}=\langle X_{\bf q}^{\sigma 0}X_{\bf q}^{0\sigma}\rangle,
\label{A9}
\end{equation}
and
%with $\gamma_{\bf q}=1/2 [\cos q_{x}+\cos q_{y}]$ and
%$\gamma_{\bf q}^{\prime}=\cos q_{x}\cos q_{y}$ for 2D square lattice
%(the lattice constant is taken to be unity).
\begin{equation}
\chi_{1}=\frac{1}{N}\sum_{\bf q}
\gamma_{\bf q}\langle S_{\bf q}^{+}S_{\bf q}^{-}\rangle,\;\;\;
\lambda_{\bf q}=\frac{1-\gamma_{\bf q}^{\prime}}{1-\gamma_{\bf q}}.
\label{A10}
\end{equation}

To calculate the correlation function in the
right hand-side of Eq. (\ref{A7}) we employ the decoupling scheme which
is essentially equivalent to the mode coupling approximation but for
the equal time correlation function. Due to the
unconventional commutation relations for HO's it is more
convenient to use the site representation
\begin{equation}
(i^{2}\ddot{S}_{\bf q}^{+},S_{\bf q}^{-})=\sum_{{\bf R}_{il}}
e^{-i{\bf q}{\bf R}_{il}}
(i^{2}\ddot{S}_{i}^{+},S_{l}^{-}),
\label{A11}
\end{equation}
where second derivative of $S_{i}^{+}$ reads
\begin{eqnarray}
i^{2}\ddot{S}_{i}^{+}&=&
\sum_{j,n}t_{ij}\Bigl\{t_{jn}\left[H_{ijn}+H_{nji}\right]
-t_{in}\left[H_{jin}+H_{nij}\right]
\Bigr\}\nonumber \\
&+&
\sum_{j,n}J_{ij}\Bigl\{J_{jn}\left[2P_{ijn}+\Pi_{nji}\right]
-J_{in}\left[2P_{jin}+\Pi_{nij}\right]
\Bigr\},
\label{A12}
\end{eqnarray}
with
\begin{eqnarray}
H_{ijn}&=&X_{i}^{+0}X_{j}^{+-}X_{n}^{0+}+
X_{i}^{+0}(X_{j}^{00}+X_{j}^{--})X_{n}^{0-},\nonumber \\
P_{ijn}&=&S_{i}^{z}S_{j}^{z}S_{n}^{+}-S_{i}^{z}S_{n}^{z}S_{j}^{+},\nonumber\\
\Pi_{ijn}&=&S_{i}^{+}S_{j}^{-}S_{n}^{+}-S_{j}^{+}S_{i}^{-}S_{n}^{+}.
\label{A13}
\end{eqnarray}
In obtaining (\ref{A12}) we have neglected terms proportional to
$tJ$ since they give no contribution within the
adopted approximation.

In the sum (\ref{A12}) only two  site indices can be equal. We extract
those terms and by using the multiplication
rules (\ref{m6}) replace the product of two HO's with the
same site indices by one operator. On rearranging, in the sum
there are no products
of operators having the same site indices.
Therefore in all products operators can be interchanged.
(Of course in the case of two Fermi operators one has to change
the sign of the product). Further, we substitute the properly
rearranged right hand-site of Eq. (\ref{A12}) into (\ref{A11}) and make
the following decoupling
\begin{eqnarray}
( X_{i}^{\sigma\sigma}S_{j}^{+},S_{l}^{-})
\simeq \langle X_{i}^{\sigma\sigma}\rangle(S_{j}^{+},S_{l}^{-})\;\;\;
\;\;\;\;\;\;\;\;
(i\not=j)\nonumber\\
(S_{i}^{+}S_{j}^{-}S_{n}^{+},S_{l}^{-})
\simeq \langle S_{i}^{+}S_{j}^{-}\rangle(S_{n}^{+},S_{l}^{-})+
\langle S_{n}^{+}S_{j}^{-}\rangle(S_{i}^{+},S_{l}^{-})\;\;\;
(i\not=j\not=n)
\label{A14}
\end{eqnarray}
In the above defined decoupling scheme the operators
on the same lattice site is never decoupled. Therefore
within the adopted approximation the local correlations are
retained.

In the momentum space the above defined decoupling
scheme results in the following expression:
\begin{equation}
(i^{2}\ddot{S}_{\bf q}^{+},S_{\bf q}^{-})\simeq\omega_{\bf q}^{2}
(S_{\bf q}^{+},S_{\bf q}^{-})=\omega_{\bf q}^{2}\chi_{\bf q},
\label{A15}
\end{equation}
where
\begin{equation}
\omega_{\bf q}^{2}=4J^{2}z^{2}|\chi_{1}|(1-\gamma_{\bf q})
[1+\Delta+C\lambda_{\bf q}+\gamma_{\bf q}],
\label{A16}
\end{equation}
with the following notations:
\begin{equation}
\Delta=\frac{1}{|\chi_{1}|}\Bigl\{\chi_{2}+\frac{1-z}{z}|\chi_{1}|
+\alpha-\eta\Bigr\},\;\;\;
C=\frac{\alpha^{\prime}-\eta}{|\chi_{1}|},
\label{A17}
\end{equation}
\begin{equation}
\chi_{2}=\frac{1}{N}\sum_{\bf q}\gamma_{\bf q}^{2}
\langle S_{\bf q}^{+}S_{\bf q}^{-}\rangle,
\label{A18}
\end{equation}
\begin{equation}
\alpha=\frac{t^{2}}{2NJ^{2}}\sum_{\bf q}\gamma_{\bf q}^{2}
\tilde{n}_{\bf q},\;\;
\alpha^{\prime}=\frac{(t^{\prime})^{2}}{2NJ^{2}}\sum_{\bf q}
(\gamma_{\bf q}^{\prime})^{2}\tilde{n}_{\bf q},
\;\;
\eta=\frac{tt^{\prime}}{2NJ^{2}}\sum_{\bf q}
\gamma_{\bf q}\gamma_{\bf q}^{\prime}n_{\bf q},
\label{A19}
\end{equation}
and
\begin{equation}
\tilde{n}_{\bf q}=(1-\frac{n}{2}-n_{\bf q}),\;\;\;
n=\langle X_{i}^{\sigma\sigma}\rangle.
\label{A20}
\end{equation}

Therefore, from Eq. (\ref{A7}) by using (\ref{A8}) and (\ref{A15})
for the static spin susceptibility
we obtain the following representation
\begin{equation}
\chi_{\bf q}=\frac{\langle [i\dot{S}_{\bf q}^{+},S_{\bf q}^{-}]\rangle}
{\omega_{\bf q}^{2}}=\frac{m_{\bf q}}{\omega_{\bf q}^{2}}.
\label{A21}
\end{equation}

Essentially, the equation for static susceptibility  (\ref{A21}) with
 expressions for $m_{\bf q}$  (\ref{A8}) and $\omega_{\bf q}$
(\ref{A16}) coincides with that one obtained in Ref.\cite{shi}
and can be evaluated self-consistently from the one-particle GF.

%%%%%%%%%%%%%%%%%%%%%%%%%%%%%%%%%%%%%%%%%%%%%%%%%%%%%%%%%%%%%%%

\end{document}